\documentstyle[12pt]{article}

\newcommand{\bb}{\begin{equation}}
\newcommand{\ee}{\end{equation}}
\newcommand{\bqn}{\begin{eqnarray}}
\newcommand{\eqn}{\end{eqnarray}}

\input{tcilatex}
\begin{document}

\title{A note on the spin connection representation of gravity}
\author{Mauricio Contreras$^{1}$ and Jorge Zanelli.$^{1,2}$ \\
$^{1}$Centro de Estudios Cient\'{\i }ficos de Santiago, Casilla 16443,
Santiago, Chile.\\
$^{2}$Universidad de Santiago de Chile, Casilla 307, Santiago 2, Chile.}
\date{}
\maketitle

\begin{abstract}
The formulation of gravity in $3+1$ dimensions in which the spin connection
is the basic field ($\omega $-frame) leads to a theory with first and second
class constraints. Here, the Dirac brackets for the second class constraints
are evaluated and the Dirac algebra of first class constraints is found to
be the usual algebra associated to space-time reparametrizations and tangent
space rotations. This establishes the classical equivalence with the
vierbein approach ($e$-frame). The explicit form of the path integral for
this theory is given and the quantum equivalence with the $e$-frame is also
established.
\end{abstract}

\section{Introduction}


The standard description of the spacetime geometry in General Relativity
uses the metric tensor $g_{\mu \nu }$ as the fundamental field. In
hamiltonian form, the action is a functional of the spatial metric $h_{ij}$
and its canonical momentum $\pi ^{ij}$, as well as four Lagrange multipliers
associated with spatial reparametrizations, $N^{i}$ (shifts) and normal
deformations, $N^{\bot \text{ }}$(lapse) \cite{ADM,Claudio}\footnote{%
In this analysis the torsion tensor is assumed to vanish identically.}.

The Einstein-Hilbert action can also be written in terms of the spin
connection $\omega _{\mu }^{ab}$, the tetrad field $e_{\nu }^{a}$ and their
exterior derivatives (first order formalism)\cite{zumino,regge}. In this
form, the hamiltonian construction needed to identify the dynamical degrees
of freedom is not straightforward. The torsion tensor does not vanish
identically, but only as a consequence of the classical equations. Thus, it
is not legitimate to eliminate the spin connection for the vierbein and one
is left with a theory for 40 independent fields $\omega _{i}^{ab}$ (18), and 
$e_{i}^{a}$ (12). Also, many of the fields do not have time derivatives: out
of the 40, only 12 have time derivatives in the lagrangian. This gives rise
to a number of first and second class constraints.


\subsection{The two frames}


An additional difficulty is that there are two natural choices of
coordinates and momenta, which have radically different phase space and
constraint structure. Thus, two options arise, depending on whether the
lagrangian involves only time derivatives of the vierbein ($e$-frame), or
the spin connection ($\omega$-frame). These two choices are related by a
canonical transformation (they differ by a total derivative) and therefore
should be classically identical in content\footnote{%
It is extremely difficult to establish the quantum mechanical equivalence
between canonically related formulations of a theory, and the equivalence
may not even exist. It has been shown that quantum mechanics could be
formulated in a way that is invariant under the simpler class of point
canonical transformations \cite{JL}, but a similar proof for general
canonical transformations is not yet known.}. However, as the corresponding
phase spaces are so radically different, proving the equivalence even at the
classical level is non trivial.

To make the discussion more concrete, let us recall some facts about the
first order formulation of gravity. The first order Lagrangian is the
Einstein-Hilbert four-form (wedge product of forms is understood) 
\begin{equation}
L_{E-H}=\epsilon _{abcd}R^{ab}e^{c}e^{d}+dB,  \label{EH}
\end{equation}
where $R^{ab}=d\omega ^{ab}+\omega _{c}^{a}\omega ^{cb}$ is the curvature
two-form, $\omega $ is the spin connection one-form, $e$ is the vierbein
one-form and $dB$ stands for some arbitrary boundary term. Different choices
of $B(e,\omega )$ give rise to different choices of canonical coordinates
(frames). Two natural choices are: 

\subsubsection{\bf The e-frame \newline
}

In the hamiltonian analysis of this action in first order form \cite{BC} the
spin connection splits in two pieces. One of them corresponds to the
canonical momentum of the tetrad field, and the other corresponds to a set
of auxiliary variables that can be eliminated from their own equations of
motion in terms of the tetrad. The resulting hamiltonian is a functional of
the tetrad and its canonical momentum only, and the spin connection drops
out. In this way the usual vierbein formulation of gravity is obtained \cite
{marc}. 

\subsubsection{{\bf The }$\omega ${\bf -frame\newline
}}

Alternatively, one can start from the Einstein-Hilbert action, eliminate the
tetrad field and build a hamiltonian action that depends on the spin
connection and its canonical momentum only \cite{Peldan}. A preliminary
discussion of the equivalence between the $\omega $ and $e$-frames was
presented in \cite{Contreras}. In this letter we want address some points of
the analysis in the $\omega $-frame. 

\subsubsection{\bf The Ashtekar approach\newline
}

The alternative approach to canonical gravity proposed by Ashtekar \cite
{Ashtekar} in the past decade is yet another canonically equivalent
description of General Relativity. The Ashtekar frame is obtained through a
complex canonical transformation from the $e$-frame \cite{HNS}. It has been
often discussed whether Ashtekar's theory is quantum mechanically equivalent
to standard metric gravity and the answer still seems uncertain and possibly
irrelevant. As we show here, the $\omega $ and $e$-frames are not only
equivalent classically through a real canonical transformation but, if there
were a quantum description for either one, it would be equivalent to the
quantum description for the other.


\section{First order formalism (in the $\omega$ frame)}


As shown in \cite{Peldan,Contreras}, dropping the boundary term in (\ref{EH}%
), the first order action for gravity in 3+1 dimensions can also be written
as\footnote{%
Our conventions are that $\epsilon ^{ijk}=\pm 1,0$ is a tensor density of
weight 1 (i.e., it transforms like a tensor of third rank times $\sqrt{g}$).
Hence, $\epsilon _{ijk}\equiv g_{il}g_{jm}g_{kn}\epsilon ^{lmn}$ is also a
tensor density of weight 1, but it takes values $\pm g,0$.} 
\begin{equation}
I[w,e]=\int (\dot{\omega}_{k}^{ab}\epsilon _{abcd}\epsilon
^{ijk}e_{i}^{c}e_{j}^{d}+\omega _{0}^{ab}J_{ab}+e_{0}^{a}P_{a}),
\label{first}
\end{equation}
%
where $J_{ab}=\epsilon _{abcd}\epsilon
^{ijk}T_{ij}^{c}e_{k}^{d}=D_{k}(\epsilon _{abcd}\epsilon
^{ijk}e_{i}^{c}e_{j}^{d})$, $P_{a}=(\epsilon _{abcd}\epsilon
^{ijk}R_{ij}^{bc}e_{k}^{d})$. If $e_{0}^{a}$ is descomposed as $%
e_{0}^{a}=N\eta ^{a}+N^{i}e_{i}^{a}$, where $\eta ^{a}$ is the normal to
espacelike surfaces, $\eta _{a}e_{i}^{a}=0$, the action can be written in
terms of the $\omega _{i}^{ab}$ and its canonically conjugate momentum $%
P_{ab}^{k}=\epsilon _{abcd}\epsilon ^{ijk}e_{i}^{c}e_{j}^{d}$, in the form 
\begin{equation}
I=\int (\dot{\omega}_{k}^{ab}P_{ab}^{k}-\omega _{0}^{ab}J_{ab}+NH_{\perp
}+N^{i}H_{i}+\mu _{ij}\phi ^{ij}),
\end{equation}
where 
\begin{equation}
H_{\perp }=g^{-1/2}P_{ac}^{i}P_{b}^{cj}R_{ij}^{ab},
\end{equation}
\begin{equation}
H_{i}=P_{ab}^{j}R_{ij}^{ab}-\omega _{i}^{ab}J_{ab},
\end{equation}
\begin{equation}
J_{ab}=D_{i}P_{ab}^{i},
\end{equation}
and 
\begin{equation}
\phi ^{ij}=\epsilon ^{abcd}P_{ab}^{i}P_{cd}^{j}.
\end{equation}
Here $\omega _{0}^{ab}$, $N$, and $N^{i}$ are Lagrange multipliers
corresponding to the constraints $J_{ab}=H_{\perp }=H_{i}=0$, and $%
g=det(g_{ij})$, $g_{ij}\equiv e_{i}^{a}e_{aj}$.

The presence of the constraint $\phi^{ij}=0$ deserves some discussion. The
substitution of $P_{a b}^k$ for $\epsilon^{i j k} \epsilon_{a b c d} e^c_i
e^d_j $ conceals the fact that there are only 12 independent fields ($e^a_i$%
) and not 18 ($P_{a b}^i$) in the phase space. The elimination of the 6
spurious fields is enforced by the 6 conditions $\phi^{i j} = 0$. The
Jacobian of the transformation $e^a_i \rightarrow P_{a b}^i$ is $\Omega_{a b
\ c}^{i \ \ j} = 2 \epsilon_{a b c d} \epsilon^{i j k} e^d_k$, which has
maximun rank (twelve) on configurations for which the local orthonormal
frames $e^a_i$ are generic, that is, they span a 3-dimensional volume (see
below).

Once the second class constraints have been eliminated, $H_i$ and $J_{a b}$
become the generators of spatial diffemorphism and local rotations,
respectively.

Preservation in time of the constraint $\phi ^{ij}=0$ implies a new
constraint 
\begin{equation}
\chi ^{kl}\delta (x,y)=\{\phi ^{kl}(x),H_{\perp
}(y)\}=g^{-1/2}D_{i}(P_{ec}^{(k})P_{ab}^{l)}P_{f}^{ci}\epsilon ^{abfe}\delta
(x,y),  \label{chi}
\end{equation}
where the parentheses indicate symmetrization in $k$, $l$. Preservation of $%
\chi ^{kl}=0$ in turn, implies 
\begin{equation}
N\{H_{\perp },\chi ^{kl}\}+\mu _{mn}\{\phi ^{mn},\chi ^{kl}\}=0.
\label{chidot}
\end{equation}
These are equation for the Lagrange multipliers, which can be solved for $%
\mu $ because the constraints $\phi ^{mn}=\chi ^{kl}=0$ obey a second class
algebra, 
\begin{equation}
\begin{array}{ll}
\{\phi ^{mn}(x),\phi ^{kl}(y)\} & =0\nonumber \\ 
\{\chi ^{mn}(x),\chi ^{kl}(y)\} & \neq 0\nonumber \\ 
\{\chi ^{ij}(x),\phi ^{mn}(y)\} & =g^{-1/2}B^{ijmn}(x,y) \\ 
& =g^{-1/2}(G^{ijmn}(\tilde{g}(x))+G^{ijmn}(\phi (x)))\delta (x-y).\label%
{second}
\end{array}
\end{equation}
where $G^{ijmn}(A)$ is the inverse supermetric for a symmetric matrix $%
A^{ij} $ \footnote{%
Here, the metric is not defined yet, $\tilde{g}^{ij}$ is just a shorthand
for $P_{ab}^{i}P^{abj}$ which will eventually be related to the canonical
metric through $\tilde{g}^{ij}=-8gg^{ij}$.} 
\begin{equation}
G^{ijkl}(A)=2A^{ij}A^{kl}-A^{il}A^{kj}-A^{ik}A^{lj}.
\end{equation}
\newline
The precise form of $\{\chi ,\chi \}$ is not essential as we will see below.
The matrix $B^{ijmn}(x,y)$ has a formal inverse $B_{ijmn}$ given by the
series 
\begin{equation}
B=G(g)-G(g)G^{-1}(\phi )G(g)+G(g)G^{-1}(\phi )G(g)G^{-1}(\phi )G(g)+...
\end{equation}
where we have defined $G\sim G_{ijkl}$, $B\sim B_{ijkl}$, $G^{-1}\sim
G^{ijkl}$, etc.

Obviously $B_{ijmn}$ coincides with the supermetric $G_{ijmn}$ on the
constraint surface $\phi =0$, $\chi =0$. In this way we can solve (\ref
{chidot}) for $\mu _{ij}$, 
\begin{equation}
\mu _{ij}=\frac{1}{2}B_{ijmn}N\{H_{\perp },\chi ^{mn}\}.
\end{equation}
Thus no new constraints appear from the preservation in time of $\chi $. The
initial $H_{\perp }$ has nonvanishing Poisson brackets with $\phi $ or $\chi 
$, but the modified one 
\begin{equation}
\tilde{H}_{\perp }=H_{\perp }-\frac{1}{N}\mu _{ij}\phi ^{ij},
\end{equation}
does.


\section{Dirac brackets}


The second class constraints can be eliminated through Dirac bracket \cite
{Dirac,HT} defined by 
\begin{equation}
\{ U, V \}^{*} = \{ U, V \} - \{ U, \varphi^{\alpha} \} C_{\alpha \beta} \{
\varphi^{\beta}, V \} ,  \label{cordi}
\end{equation}
where $C_{\alpha \beta}$ is the inverse of the Dirac matrix $C^{\alpha
\beta} = \{ \varphi^{\alpha} , \varphi^{\beta} \}$, where $\varphi^{\alpha}$
denote generic second class constraints. In our case, the Dirac matrix 
\begin{equation}
C^{\alpha \beta} (x,y) = \left( 
\begin{array}{cc}
\{ \chi^{i j}(x) , \chi^{m n}(y) \} & \{ \chi^{i j}(x) , \phi^{m n}(y) \} \\ 
\{ \phi^{i j}(x) , \chi^{m n}(y) \} & \{ \phi^{i j}(x) , \phi^{m n}(y) \}
\end{array}
\right) ,  \label{dmx}
\end{equation}
has the form $\left( 
\begin{array}{cc}
A & B^{-1} \\ 
- B^{-1} & 0
\end{array}
\right) $, because $\{ \phi^{i j}, \phi^{m n} \} = 0 $. \newline
Its inverse takes the form $C^{-1} = \left( 
\begin{array}{cc}
0 & - B \\ 
B & B A B
\end{array}
\right) $, or 
\begin{equation}
C^{-1}_{\alpha \beta} = \left( 
\begin{array}{cc}
0 & - B_{i j m n} \\ 
B_{i j m n} & B_{i j p q} \{ \chi^{p q} , \chi^{k l} \} B_{k l m n}
\end{array}
\right) .
\end{equation}
The Dirac bracket for two arbitry functionals U, V is given by: 
\begin{equation}
\begin{array}{ll}
\{ U , V \}^{*} = & \{ U , V \} - \{ U , \chi \} B \{ \phi, V \} \\ 
& - \{ U , \phi \} B \{ \chi, V \} - \{ U , \phi \} B \{ \phi, \phi \} B \{
\phi, V \} ,
\end{array}
\label{abd}
\end{equation}
where sum and integration over discrete and continuous indices is assumed. 
\newline
It can be shown that when U and V belong to the set $\{\tilde{H}_{\perp} ,
H_i , J_{a b} \} $, the second term of the right hand side of (\ref{abd})
vanishes on the constraint surface $\phi = 0$, $\chi = 0$. In particular,
direct subtitution in (\ref{abd}) yields 
\begin{equation}
\{ \tilde{H}_{\perp}, \tilde{H}_{\perp}\}^* = \{H_{\perp}, H_{\perp}\}^* =
\{H_{\perp}, H_{\perp}\}.
\end{equation}
and using the results of \cite{Peldan}, we finally have 
\begin{equation}
\{ H_{\perp}, H_{\perp} \}^{*} \sim g^{i j} H_j \partial_i \delta(x,y) .
\end{equation}
In the same way, the complete Dirac algebra can be shown to be given by 
\begin{equation}
\{ H_{\perp}[N] , H_{\perp}[M] \}^{*} = \int [ (\partial_i N) M -
(\partial_i M) N ] g^{i j}(P) H_j ,  \label{h-h}
\end{equation}
\begin{equation}
\{ H_{\perp} [N] , H_i [M^{i}] \}^{*} = \int ( M^{i} \partial_{i} N - N
\partial_{i} M^{i} ) H_{\perp} ,
\end{equation}
\begin{equation}
\{ H_i [M^{i}] , H_j [M^{j}] \}^{*} = \int (N^{l} \partial_{l} M^{m} - M^{l}
\partial_{l} N^{m}) H_{m} ,
\end{equation}
\begin{equation}
\{ J[N^{a b}] , J[M^{c d}] \}^{*} = \int J[(M \times N)^{a b}] ,
\end{equation}
\begin{equation}
\{ J[N^{a b}], H[M] \}^{*} = 0 ,
\end{equation}
\begin{equation}
\{ J[N^{a b}], H[M^{i}] \}^{*} = 0 .  \label{j-h}
\end{equation}
Thus, the Dirac algebra reduces to a direct sum of the usual algebra of
spacetime diffeomorphism plus tangent space rotations.

Note that when $P_{a b}^k$ is replaced by its expresion in terms of the
tetrad, the $\phi^{i j}$ constraints vanish identically, but the secondary
constraints $\chi^{i j}$ do not. In the vierbein frame \cite{BC} it can also
be shown that prior to eliminating the auxiliary variables, apart from $J_{a
b}, H_{\perp}, H_{i}$ the constraints 
\begin{equation}
\gamma^{i j} = E_{a}^{(i} \ \epsilon^{m n j)} \ T^{a}_{m n} = 0 ,
\label{tilphi}
\end{equation}
are found, where $T^{a}_{i j}$ are the spatial components of the torsion
tensor, and $E_a^i \equiv e_{a j} \ g^{i j}$. Equation (\ref{tilphi}) is one
of the field equations, from which the auxiliary variables $\lambda^{i j}$
can be eliminated. Replacing $P_{a b}^i = \epsilon^{ijk} \epsilon_{a b c d}
e^c_j e^d_k $ in the definition (\ref{chi}), $\chi^{i j}$ can be identified
with $\gamma^{i j}$ in the e-frame.

The algebra (\ref{h-h}--\ref{j-h}) is the same as the one found in the
vielbein-frame once the contraint $\gamma^{ij}$ is strongly set equal to
zero. The two frames can be compared and contrasted in the following table: 
\newline
\newline
\newline
\begin{tabular}{|c|c|c|}
\hline
\ \ \  & e-frame & $\omega$-frame \\ \hline
&  &  \\ 
dynamical & $e^a_i, \pi_a^i = \epsilon^{ijk} \epsilon_{a b c d} e^b_j
\omega^{c d}_k $ & $\omega^{a b}_i, P_{a b}^i = \epsilon^{ijk} \epsilon_{a b
c d} e^c_j e^d_k $ \\ 
variables &  &  \\ 
(q,p) & (12) , (12) & (18) , (18) \\ 
&  &  \\ \hline
First class &  &  \\ 
constraint & $H_{\perp}, H_{i}, J_{a b}$ & $H_{\perp}, H_{i}, J_{a b}$ \\ 
&  &  \\ \hline
second class &  &  \\ 
constraint & --------- & $\chi^{ij}, \phi^{kl}$ \\ 
&  &  \\ \hline
prop. &  &  \\ 
degrees of & 12 - 4 - 6 = 2 & 18 - 10 - $\frac{1}{2}$ 12 = 2 \\ 
freedom &  &  \\ 
&  &  \\ \hline
\end{tabular}
\newline
\newline

(Here $\gamma^{ij}$ has been eliminated in the $e$-frame). The number of
propagating degrees of freedom is $g = c - f - \frac{1}{2}s $, where $c$ is
the number of coordinates, $f$ the number of first class constraints and $s$
the number of second class constraints.


\section{Path integral}


We now consider the path integral for this system. As shown in \cite{HT},
the path integral for a system with second class constraints $\chi$, $\phi$
has a measure proportional to 
\begin{equation}
\delta (\chi) \delta(\phi) \sqrt{det C^{\alpha \beta}} ,  \label{measure}
\end{equation}
where $C^{\alpha \beta}$ is the Dirac matrix. In our case, $C^{\alpha \beta}$
as given in (\ref{dmx}) and (\ref{second}), yields 
\begin{equation}
\sqrt{det C^{\alpha \beta} } = det( G^{i j m n}(\tilde{g})+G^{i j m n}(\phi)
) .
\end{equation}
The delta functions in (\ref{measure}) restrict the integration to the
constraint surface $\phi = 0$, $\chi = 0$, so path integral reads 
\begin{equation}
Z = \int [DP_{a b}^{k}] [D\omega^{a b}_{k}] det(G^{ijmn}(\tilde{g}))
\delta(\phi^{ij}) \delta(\chi^{mn}) det M_{\alpha \beta} \delta(H_{\perp})
\delta(H_{i}) \delta(J_{a b}) \ exp \frac{i}{\hbar} S  \label{zeta}
\end{equation}
with 
\begin{equation}
S = \int \dot{\omega}^{a b}_{k} P_{a b}^{k} ,
\end{equation}
and $M_{\alpha \beta}$ is the matrix of Poisson brackets 
\begin{equation}
M_{\alpha \beta} = \{ F_{\alpha} , \varphi_{\beta} \}^* ,
\end{equation}
where $F_{\alpha}$ are gauge condition for the first class constraint set $%
\varphi_{\beta} = \{ H_{\perp}, H_i , J_{a b} \}$.


\section{{\bf The }$\omega ${\bf -}$e${\bf \ transformation}}

Consider now the following transformation, wich maps the 18 coordinates $%
\omega _{i}^{ab}$ and their 18 canonically conjugate momenta $P_{ab}^{i}$
into 12 $e_{i}^{a}$'s, 12 $\pi _{a}^{i}$'s, 6 auxiliary variables $\lambda
_{mn}$ and 6 $\rho ^{mn}$ ( $\lambda _{mn}$ and $\rho ^{mn}$ are symmetric
and m,n take the values 1,2,3.) 
\begin{equation}
\omega _{k}^{ab}=\Theta _{k\ \ j}^{ab\ c}\ \pi _{c}^{j}+U_{k}^{ab\ \ mn}\
\lambda _{mn}  \label{trans1}
\end{equation}
\begin{equation}
P_{ab}^{k}=\frac{1}{2}\Omega _{ab\ c}^{k\ \ j}\ e_{j}^{c}+V_{ab\ \ mn}^{k}\
\rho ^{mn}.  \label{trans2}
\end{equation}
Here $\Theta $ and $\Omega $ are rectangular matrices, 
\begin{equation}
\Theta _{i\ \ j}^{ab\ c}=\frac{1}{8\sqrt{g}}[e_{i}^{[a}\eta
^{b]}e_{j}^{c}-e_{i}^{[a}e_{j}^{b]}\eta ^{c}-2e_{j}^{[a}\eta ^{b]}e_{i}^{c}],
\end{equation}
\begin{equation}
\Omega _{ab\ c}^{i\ \ j}=2\epsilon _{abcd}\ e_{k}^{d}\ \epsilon ^{ijk},
\end{equation}
where the square brackets indicate antisymmetrization. $U$ and $V$ are null
vectors for $\Omega $ and $\Theta $, given by 
\begin{equation}
U_{k}^{ab\ \ mn}=\frac{1}{2}\delta _{i}^{(m}\ \epsilon ^{n)kl}\ e_{k}^{a}\
e_{l}^{b},
\end{equation}
\begin{equation}
V_{ab\ \ mn}^{k}=\frac{1}{g}\ E_{a}^{r}\ E_{b}^{s}\epsilon _{rs(m}\ \delta
_{n)}^{i}.
\end{equation}
These objects satisfy the following relations 
\begin{equation}
\Omega _{ab\ c}^{k\ \ i}\ \Theta _{k\ \ j}^{ab\ d}=\delta _{d}^{c}\delta
_{j}^{i},  \label{orto1}
\end{equation}
\begin{equation}
\Omega _{ab\ c}^{k\ \ j}\ U_{k}^{ab\ \ mn}=0,
\end{equation}
\begin{equation}
\Theta _{k\ \ j}^{ab\ c}\ V_{ab\ \ mn}^{k}=0,
\end{equation}
\begin{equation}
U_{k}^{ab\ \ mn}\ V_{ab\ \ pq}^{k}=\delta _{(pq)}^{(mn)}.  \label{orto4}
\end{equation}
One can think of $\Theta $ and $\Omega $ as a collection of twelve vectors
--labeled by the indices $(_{i}^{a})$ and $(_{a}^{i})$ respectively--, in an
18-dimensional vector space with components $(_{j}^{ab})$, and $(_{ab}^{j})$%
, respectively. By the same token, $U$ and $V$ are six vectors (labeled by
the index $(mn)$) in an 18-dimensional vector space.

In this sense the properties (\ref{orto1},...\ref{orto4}) are nothing but
orthogonality relations among the vectors $\Theta$, $\Omega $, $U$ and $V$.
These relations imply the following completeness relation 
\begin{equation}
\Theta^{ab \ e}_{i \ \ l} \ \Omega_{cd \ e}^{j \ \ l} + U^{ab \ \ mn}_{i} \
V_{cd \ \ mn}^{j} = \delta^{[a b]}_{[c d]} \delta^i_j ,  \label{completeness}
\end{equation}
which will be used in what follows. In this way, the 18 vectors $\Theta^{ab
\ c}_{i \ \ j}$, $U^{ab \ \ mn}_{i}$ are a basis of the space of
contravariant vectors $L^{ab}_i$, while $\Omega_{ab \ c}^{i \ \ j}$, $V_{ab
\ \ mn}^{j}$ are a basis for the dual (covariant vectors, $L_{ab}^i$). Thus,
the field transformations (\ref{trans1}), (\ref{trans2}) correspond to the
expansions of $\omega^{ab}_i$ and $P_{ab}^i$ in the contravariant and
covariant bases, respectively.

As shown in the appendix, using (\ref{trans1}), (\ref{trans2}) the path
integral (\ref{zeta}) can now be written in terms of the coordinates of the $%
e$-frame as 
\begin{equation}
Z=\int [De_{i}^{a}][D\pi _{c}^{j}]\ detM_{\alpha \beta }\ \delta (H_{\perp
})\delta (H_{i})\delta (J_{ab})\ exp\frac{i}{\hbar }S,
\end{equation}
which is the path integral one would write in the $e$-frame. This shows the
equivalence between quantum theories one would obtain in the two frames. The
different constraints $H_{\perp }$, $H_{i}$,and $J_{ab}$ can be written
explicitly in terms of $e$-frame variables, as 
\begin{equation}
\frac{1}{2}H_{\perp }=\eta ^{a}\partial _{i}\pi _{a}^{i}-\frac{1}{2}%
E_{d}^{s}\partial _{[l}e_{s]}^{d}\eta ^{b}\pi _{b}^{l}-G_{\perp ij}^{ab}\pi
_{a}^{i}\pi _{b}^{j}-g^{3/2}G^{mnpq}\lambda _{mn}^{0}\lambda _{pq}^{0},
\end{equation}
\begin{equation}
\begin{array}{ll}
N^{m}H_{m}= & N^{m}[\frac{1}{2}(g^{-1}E_{d}^{s}\partial
_{i}e_{k}^{d}\epsilon _{mls}\epsilon ^{ijk}e_{j}^{b}-E_{d}^{s}\partial
_{[m}e_{s]}^{d}e_{l}^{b}+\eta _{d}\partial _{[m}e_{l]}^{d}\eta ^{b})\pi
_{b}^{l} \\ 
& +e_{m}^{a}\partial _{i}\pi _{a}^{i}+G_{mij}^{ab}\pi _{a}^{i}\pi _{b}^{j}]+%
\frac{1}{2}N^{(m}e_{i}^{a}e_{j}^{b}J_{ab}\epsilon ^{ijn)}\lambda _{mn}^{0}
\\ 
& -N^{i}\omega _{i}^{ab}\ J_{ab},
\end{array}
\end{equation}
\begin{equation}
J_{ab}=2\epsilon _{abcd}\frac{\partial e_{j}^{c}}{\partial x^{i}}%
e_{k}^{d}\epsilon ^{ijk}-\frac{1}{2}(\pi _{a}^{i}e_{bi}-\pi _{b}^{i}e_{ai}),
\end{equation}
where 
\begin{equation}
\lambda _{pq}^{0}=\frac{1}{2g}G_{pqmn}E_{a}^{(m}\partial
_{i}e_{j}^{a}\epsilon ^{ijn)},  \label{lambda0}
\end{equation}
\begin{equation}
G_{\perp ij}^{ab}=\frac{1}{16\sqrt{g}}[%
e_{i}^{a}e_{j}^{b}-2e_{j}^{a}e_{i}^{b}-g_{ij}\eta ^{a}\eta ^{b}],
\end{equation}
and 
\begin{equation}
G_{mij}^{ab}=\frac{1}{16\sqrt{g}}[g_{ij}\eta
^{a}e_{m}^{b}+2g_{im}(e_{j}^{a}\eta ^{b}-e_{j}^{b}\eta ^{a})].
\end{equation}
Finally, the kinetic term $P_{ab}^{i}\dot{\omega}_{i}^{ab}$ in the action S
reduces via the $e$-$\omega $ transformation to the usual $e$-frame kinetic
term $\pi _{a}^{i}\dot{e}_{i}^{a}$. This completes the classical and quantum
equivalence between the $\omega $ and $e$ frames.


\begin{quotation}
{\LARGE Acknowledgments}
\end{quotation}

The authors are grateful to M. Ba\~{n}ados, M. Henneaux and R. Troncoso for
many enlightening discussions and helpful comments. This work was supported
in part through grants 1960229, 1970151, 1980788 3960007, and 7960001 from
FONDECYT, and grant 27-953/ZI-DICYT (USACH). The institutional support of
FUERZA AEREA DE CHILE, I.\ Municipalidad de Las Condes, and a group of
Chilean companies (AFP Provida, Business Design Associates, CGE, CODELCO,
COPEC, Empresas CMPC, GENER\ S.A., Minera Collahuasi, Minera Escondida,
NOVAGAS and XEROX-Chile) is also recognized.

\vspace{1cm}


{\bf Appendix: Equivalence of the measure in the $\omega $ and $e$ frames} 
\newline

Using (\ref{trans1},\ref{trans2}), the path integral in the $\omega $ frame
given in (\ref{zeta}) reads 
\begin{equation}
\begin{array}{ll}
Z=\int & [De_{i}^{a}][D\pi _{a}^{i}][D\lambda _{mn}][D\rho ^{mn}]\ J\
det(g^{3/2}G^{ijkl}(\tilde{g}))\ \delta (\phi ^{ij})\delta (\chi ^{mn}) \\ 
& detM_{\alpha \beta }\ \delta (H_{\perp })\delta (H_{i})\delta (J_{ab})\ exp%
\frac{i}{\hbar }S
\end{array}
\end{equation}
where $J$ is the determinant of the Jacobian matrix of the transformation $%
(\omega ,P)\rightarrow (e,\pi ,\lambda ,\rho )$.

The different constraints must be written in terms of the new variables.
Consider first $\phi ^{ij}=\epsilon ^{abcd}P_{ab}^{i}P_{cd}^{j}$. Using (\ref
{trans2}), $\phi ^{ij}$ it is easily shown that 
\begin{equation}
\phi ^{ij}=32\rho ^{ij},
\end{equation}
so that $\delta (\phi ^{ij})=\delta (32\rho ^{ij})$. In the same way,
substituting (\ref{trans2}) in (\ref{chi}), the constraints $\chi ^{ij}$
become 
\begin{equation}
\chi ^{ij}=\frac{g^{1/2}}{2}E_{a}^{(i}\ \epsilon ^{j)mn}\ T_{mn}^{a}
\end{equation}
which are recognized as the second class constraints in the $e$-frame (\ref
{tilphi}). The $\chi $ constraints can be rewritten substituting $\omega $
from (\ref{trans1}) in $T_{ij}^{a}(\omega ,e)$ in the form 
\begin{equation}
\chi ^{ij}=-2g^{3/2}G^{ijmn}(g)(\lambda _{mn}-\lambda _{mn}^{0}),
\end{equation}
where $\lambda _{mn}^{0}$ are given by (\ref{lambda0}), then 
\begin{eqnarray}
\delta (\chi ^{ij}) &=&\delta (-2g^{3/2}G^{ijmn}(g)(\lambda _{mn}-\lambda
_{mn}^{0})) \\
&=&\frac{\delta (\lambda _{mn}-\lambda _{mn}^{0})}{det(-2g^{3/2}G^{ijmn}(g))}%
.  \nonumber
\end{eqnarray}
The metric $\tilde{g}^{ij}=P_{ab}^{i}P^{abj}$ becomes, after using (\ref
{trans2}), $\tilde{g}^{ij}=-8gg^{ij}$, so $G^{ijmn}(\tilde{g}%
)=64g^{2}G^{ijmn}(g)$ . \newline
Thus, the path integral reads, up to a normalization constant, 
\begin{equation}
\begin{array}{ll}
\begin{array}{l}
\\ 
Z[e,\pi ,\lambda ,\rho ]= \\ 
\end{array}
\int  & [De_{i}^{a}][D\pi _{a}^{i}][D\lambda _{mn}][D\rho ^{mn}]\ J\ \delta
(\rho ^{mn})\delta (\lambda _{mn}-\lambda _{mn}^{0}) \\ 
& \times \,\,det\,M_{\alpha \beta }\ \delta (H_{\perp })\delta (H_{i})\delta
(J_{ab})\ exp\frac{i}{\hbar }S.
\end{array}
\end{equation}
Integrating over $\lambda $ and $\rho $ one obtains 
\begin{equation}
Z[e,\pi ]=\int [De_{i}^{a}][D\pi _{c}^{j}]\ J_{0}\ detM_{\alpha \beta }\
\delta (H_{\perp })\delta (H_{i})\delta (J_{ab})\ exp\frac{i}{\hbar }S,
\end{equation}
where $J_{0}$ is the Jacobian evaluated at $\lambda =\lambda _{0}$ and $\rho
=0$. Now we will show that this Jacobian is one, that is, the measure is
invariant under the transformation (\ref{trans1}), (\ref{trans2}). In what
follows we denote de collective indeces $(_{i}^{ab})\rightarrow A$, $%
(_{i}^{a})\rightarrow a$ and $(mn)\rightarrow \alpha $. Then, varying the
fields in the transformation (\ref{trans1}, \ref{trans2}) yields 
\begin{equation}
\delta \omega ^{A}=\frac{\partial \Theta ^{Ab}}{\partial e^{a}}\pi _{b}\
\delta e^{a}+\frac{\partial U^{A\beta }}{\partial e^{a}}\lambda _{\beta }\
\delta e^{a}+\Theta ^{Aa}\delta \pi _{a}+U^{A\alpha }\delta \lambda _{\alpha
},
\end{equation}
\begin{equation}
\delta P_{A}=\omega _{Aa}\delta e^{a}+\frac{\partial V_{A\beta }}{\partial
e^{a}}\rho ^{\beta }\ \delta e^{a}+0\ \delta \pi _{a}+V_{A\alpha }\ \delta
\rho ^{\alpha },
\end{equation}
so that the jacobian matrix is given by 
\begin{equation}
J=\left[ 
\begin{array}{cccc}
\frac{\partial \Theta ^{Ab}}{\partial e^{a}}\pi _{b}+\frac{\partial
U^{A\beta }}{\partial e^{a}}\lambda _{\beta } & 0 & \Theta ^{Aa} & 
U^{A\alpha } \\ 
\omega _{Aa}+\frac{\partial V_{A\beta }}{\partial e^{a}}\rho ^{\beta } & 
V_{A\alpha } & 0 & 0
\end{array}
\right] ,
\end{equation}
which has the block form $J=\left[ 
\begin{array}{cc}
A & B \\ 
C & 0
\end{array}
\right] $, so that $det(J)=det(C)\ det(B)=det(C)det(B^{t})=det(CB^{t})$. In
our case 
\begin{equation}
\begin{array}{ll}
CB^{t} & =[(\omega _{Aa}+\frac{\partial V_{A\beta }}{\partial e^{a}}\rho
^{\beta })\ \ V_{A\alpha }][\Theta ^{Ba}\ \ U^{B\alpha }] \\ 
& =\omega _{Aa}\Theta ^{Ba}+V_{A\alpha }U^{B\alpha }+\frac{\partial
V_{A\beta }}{\partial e^{a}}\rho ^{\beta }\Theta ^{Ba}
\end{array}
.
\end{equation}
The first two terms reproduce exactly the completenees relation (\ref
{completeness}), so the jacobian is 
\begin{equation}
J=det(\delta _{A}^{B}+\frac{\partial V_{A\alpha }}{\partial e^{a}}\rho
^{\alpha }\Theta ^{Ba}).
\end{equation}
Finally, evaluating the jacobian on the constraint surface $\rho =0$, $%
\lambda =\lambda ^{0}$, one finds $J|_{\rho =0}=det(\delta _{A}^{B})=1$, and
the path integral can be  finally written as
\begin{equation}
Z=\int [De_{i}^{a}][D\pi _{c}^{j}]\ detM_{\alpha \beta }\ \delta (H_{\perp
})\delta (H_{i})\delta (J_{ab})\ exp\frac{i}{\hbar }S,
\end{equation}
which is the expected expression for the path integral in the $e$-frame.


\end{document}